\documentclass[letter,twocolumn]{jpsj2} 

\newcommand{\etal}{{\it et al.}\ }

\title{First-Principles Electronic Structure of Solid Picene}
\author{Taichi Kosugi$^{1,2}$, Takashi Miyake$^{2,4}$, Shoji Ishibashi$^2$, \\ Ryotaro Arita$^{3,4}$ and Hideo Aoki$^{1}$}

\inst{
$^1$ Department of Physics, University of Tokyo, Hongo, Tokyo 113-0033, Japan \\
$^2$ Research Institute for Computational Sciences (RICS), AIST, Umezono, Tsukuba 305-8568, Japan \\
$^3$ Department of Applied Physics, University of Tokyo, Hongo, Tokyo 113-8656, Japan \\
$^4$ Japan Science and Technology Agency, CREST, Honcho, Kawaguchi, Saitama 332-0012, Japan
}

\abst{
To explore the electronic structure of the first aromatic superconductor, potassium-doped solid picene which has been recently discovered by Mitsuhashi \etal with the transition temperatures $T_c=7 - 20$ K, 
we have obtained a first-principles electronic structure 
of solid picene as a first step toward 
the elucidation of the mechanism of the superconductivity.
The undoped crystal is found to have four conduction 
bands, which are characterized in 
terms of the maximally localized Wannier orbitals.  
We have revealed how the band structure reflects the stacked 
arrangement of molecular orbitals for both undoped and 
doped (K$_3$picene) cases, where the bands are not rigid.  
The Fermi surface for K$_3$picene 
is a curious composite of a warped two-dimensional 
surface and a three-dimensional one. 
}

\kword {picene, superconductivity, aromatic compound, first-principles calculation, electronic structure}

\begin{document}
\maketitle

{\it Introduction ---} 
While  there are mounting interests toward superconductivity 
in more versatile 
classes in recent years, fascination continues with 
superconductivity in 
$\pi$-electron (organic) systems.  This has a long history, dating back 
to superconductivity in graphite intercalation compounds (GIC's), 
first discovered~\cite{bib:1114} in 1965 in alkal-metal doped AC$_8$ (A = K, Rb, Cs) with transition temperatures $T_c < 1$ K, which was 
recently raised to $11.5$ K in 
CaC$_6$\cite{bib:1108_10-11}.
The superconductivity in organic metals was first 
found in TMTSF compounds~\cite{bib:1117_9} in 1980 with $T_c = 0.9$ K under 12 kbar.  Then comes 
the BEDT-TTF family~\cite{bib:1108_13}, 
for which $T_c = 14.2$ K  under 82 kbar was  found in 2003, 
the highest $T_c$ to date among the organic superconductors. 
Another family is the doped fullerides, 
discovered~\cite{bib:1104_1} in 1991 for K$_3$C$_{60}$ with $T_c = 18$ K at ambient pressure, 
followed by Cs$_2$RbC$_{60}$ with 
$T_c = 33$ K at ambient pressure\cite{bib:1104_4} 
and Cs$_3$C$_{60}$ with $T_c = 40$ K under 15 kbar \cite{bib:1104_8}, which is the highest $T_c$ among carbon-based materials.  
However, a new class of superconducting $\pi$-electron materials 
has not been found over a decade.

So a recent discovery of superconductivity by Mitsuhashi \etal~\cite{kubozono}  in a solid of 
aromatic molecule picene, with $T_c= 7$ and $20$ K when doped with 
potassium, is a long-awaited breakthrough.
The discovery is seminal  in that this is the first aromatic superconductor, 
so that a comparison 
with other classes of $\pi$-electron superconductors should be interesting.  A picene molecule consists of five benzene rings joined in an 
armchair manner with a higher chemical stability than pentacene with a 
straight arrangement.   
The solid picene has in fact a much larger band gap 
($\simeq 3.3$ eV)\cite{bib:1106}
than that of solid pentacene ($\simeq 1.8$ eV), which is 
why the material originally attracted attention as an organic semiconductor in e.g. organic FET's~\cite{bib:1106,bib:1107}.  
Picene has been known to crystallize in orthorhombic 
and monoclinic  forms~\cite{bib:1079}, and 
Kubozono's group discovered superconductivity in 
potassium-doped monoclinic crystalline picene.\cite{kubozono}  
The crystal comprises layers stacked in $c$ direction, where each 
layer (in $ab$ plane) has picene molecules arranged 
in a herringbone structure.  
The measured lattice parameters of the crystalline picene are 
$a=8.480, b=6.154, c=13.515$ in 
\AA \ with $\beta=90.46^\circ$~\cite{bib:1079}, where a primitive cell contains two inequivalent molecules 
A and B whose positions are related with a space group $P2_1$.  
In \cite{kubozono} 
K$_x$picene with the concentration $x$ of K atoms per picene 
is found to become a superconductor for $x \simeq 3$, and 
the Meissner effect has been detected with $H_{c2}>10^4$ Oe suggested.

Theoretically, the first thing we have to do is 
first-principles electronic structure calculations, which we have done 
in this Letter, 
with the density functional theory (DFT) for both pristine 
and doped solids of picene 
as a first step to elucidate the mechanism of superconductivity.  We have identified the main characters of the conduction bands in terms of the relevant molecular orbitals, and 
we have also constructed a downfolded 
four-band tight-binding model with 
maximally localized Wannier orbtials.  
From these we shall show how the band structure reflects the stacked 
arrangement of molecular orbitals for both undoped picene and  
K$_3$picene.  The Fermi surface for K$_3$picene 
has turned out to be a curious composite of a warped two-dimensional 
surface and a three-dimensional one, where both the 
molecular arrangements modified by doping and hybridization with 
alkali-metal atoms are relevant.

\begin{figure}[htbp]
\begin{center}
\includegraphics[keepaspectratio,width=5cm]{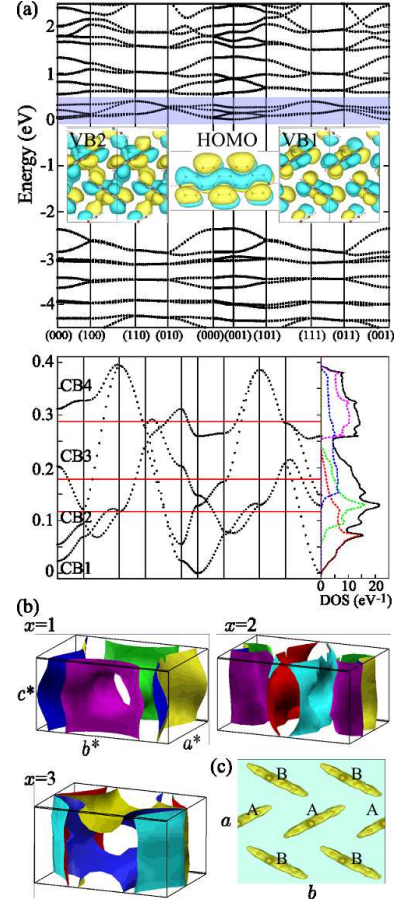}
\end{center}
\caption{
(a) Calculated band structure of undoped crystalline picene.
The horizontal axis is labeled by $(\boldsymbol{a}^*/2, \boldsymbol{b}^*/2, \boldsymbol{c}^*/2)$, while the origin of the energy is set to the conduction band bottom.
Insets show the highest occupied molecular orbital and the VB1 and VB2 Bloch wave functions at $\Gamma$.
Bottom panel is 
a blowup of the conduction bands (shaded region in the top panel), with 
the density of states per molecule shown in the solid curve
while the dashed curves the contributions from the four bands.
Horizontal lines indicate $E_{\mathrm{F}}$ for $x=1, 2$ and $3$.  
(b) The Fermi surfaces when $E_{\mathrm{F}}$ is shifted to the 
situation in K$_x$picene with $x=1, 2$ and $3$ with rigid bands.  
(c) The electron density viewed along the long molecular axis.
}
\label{band1}
\end{figure}

{\it First-principles bands ---} 
We have first optimized the atomic positions including hydrogen atoms 
with the lattice parameters fixed at the measured values~\cite{bib:1079}.  
For the electronic structure calculation for the optimized structure 
we adopt DFT
with the projector augmented-wave (PAW) method~\cite{bib:PAW} 
with the Quantum MAterials Simulator (QMAS) package~\cite{bib:QMAS} 
in the local-density approximation (LDA).  
The pseudo Bloch wave functions are expanded by plane waves 
with an energy cutoff of 40 Ry and $4^3$ $k$-points.
The angle between the planes of the inequivalent molecules in the optimized geometry is $61^\circ$, close to the measured $58^\circ$.
Each molecule is not exactly planar due to the intramolecular overcrowding, as seen in the experiment~\cite{bib:1079}, with 
the calculated center-of-mass vectors between two inequivalent, 
nearest-neighbor molecules $\boldsymbol{r}_{\mathrm{B}} - \boldsymbol{r}_{\mathrm{A}} = 0.485 \boldsymbol{a} \pm 0.5 \boldsymbol{b}$.
In the calculated electronic band structure in Fig.\ref{band1} (a), 
we first notice that the 
dispersion along the $c^*$ axis is small compared to those along 
the $a^*b^*$ plane, reflecting the layered structure.  
The lowest 102 bands are fully occupied, 
with the valence band top occurring at $\boldsymbol{c}^*/2$.  
The conduction band bottom is also located there, giving a band 
gap of 2.36 eV,
while the experimental value~\cite{bib:1106} is 3.3 eV.
Such an underestimation is generally seen in LDA calculations.  
As seen in Fig.\ref{band1} (a), the band dispersions along $k_{b^*} = \pm 1/2$ 
are doubly degenerate, unlike the triclinic crystalline tetracene and pentacene but like the monoclinic crystalline naphthalene and 
anthracene~\cite{bib:1087}.
The electron density, displayed in Fig.\ref{band1} (c), spreads over each 
molecules basically uniformly, with no significant dimerizaton seen.  

For comparison we have also calculated the electronic structures of 
an isolated picene 
molecule with the molecular geometries fixed at those optimized in the 
crystal to 
obtain the HOMO and LUMO wave functions, where 
the LUMO-HOMO energy gap is 2.96 eV (against 1.12 eV calculated 
for pentacene molecule).  
If we regard picene and pentacene as fragments of graphene with 
armchair and zigzag edges, respectively, the 
higher stability (i.e., larger gap) in picene is consistent with 
the wisdom obtained in graphene, 
where armchairs are much stabler.  

\begin{figure}[htbp]
\begin{center}
\includegraphics[keepaspectratio,width=8.5cm]{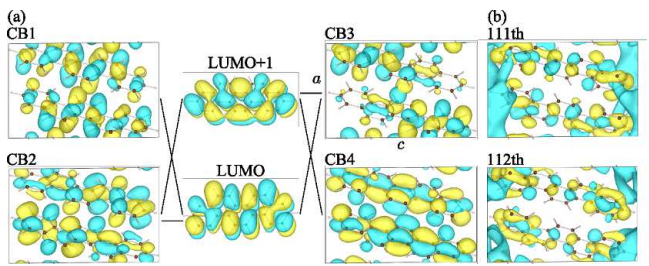}
\end{center}
\caption{
(a) The lowest unoccupied molecular orbital (LUMO) and LUMO+1 for an isolated picene molecule as compared with 
CB1-CB4 Bloch wave functions at $\Gamma$ in the crystalline picene.  
Solid lines indicate the main orbital components.
(b) 111th and 112th Bloch wave functions at $\Gamma$ 
that have amplitudes in the interlayer regions.
}
\label{orbitals2}
\end{figure}

If we go back to the crystal, valence bands are the 99th (VB4) - 102nd (VB1) bands in an energy region of width 0.77 eV 
separated from the lower ones by 0.26 eV.
Looking at the Bloch wave functions for VB1 and VB2 at $\Gamma$,
we can identify that they have mainly the character 
of the HOMO's of picene molecule.  
In each layer in the crystal, the HOMO ($\phi$) of each molecule 
splits into bonding and antibonding ones due to the 
overlap with its four adjacent inequivalent molecules to form VB1 and VB2 Bloch states, respectively.  
We can then symbolically express them, at $\Gamma$, as
$\psi_{\mathrm{VB}1} =  \phi_{\mathrm{HOMO}}^{\mathrm{A}}  -  \phi_{\mathrm{HOMO}}^{\mathrm{B}} $ and 
$\psi_{\mathrm{VB}2} =  \phi_{\mathrm{HOMO}}^{\mathrm{A}}  +  \phi_{\mathrm{HOMO}}^{\mathrm{B}} $, 
as confirmed in Fig. \ref{band1} (a).
We also note that $ \psi_{\mathrm{VB}3} $ and $ \psi_{\mathrm{VB}4} $ are 
mainly $ \phi_{\mathrm{HOMO}-1}^{\mathrm{A,B}} $ in character, 
so that all the four valence bands mainly derive 
from HOMO and HOMO$-1$.
If we now turn to the conduction bands, 
they are the 103rd (CB1) - 106th (CB4) bands with a width of 
0.39 eV, separated from the upper 
ones by 0.15 eV (Fig.\ref{band1} (a)).
A similar analysis shows that 
the Bloch states $\psi_{\mathrm{CB}1} $
- $\psi_{\mathrm{CB}4} $ mainly derive from 
$\phi_{\mathrm{LUMO}(+1)}^{\mathrm{A,B}}$ as 
summarized in Fig.\ref{orbitals2} (a).

\begin{figure}[htbp]
\begin{center}
\includegraphics[keepaspectratio,width=8.5cm]{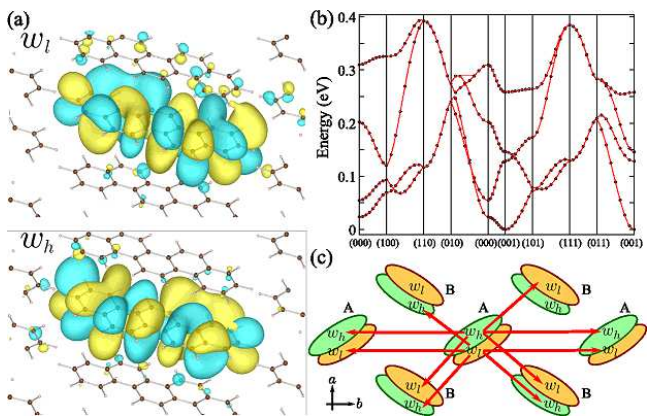}
\end{center}
\caption{
(a) Maximally localized Wannier orbitals
with the lower-energy $w_l$ and the higher-energy $w_h$,
constructed for the conduction bands.
(b) Tight-binding fit (curves) as compared with
the DFT-LDA band structure
(dots) for the four conduction bands.
(c) Arrows indicate the transfer integrals between the in-plane molecules 
whose magnitudes are larger than
$25$ meV, with ellipses representing the
associated WO's, and
overlaps representing the difference in the $c$-axis height.
}
\label{wannier}
\end{figure}

{\it Downfolding ---} 
To build a tight-binding model for the electronic structure of the pristine picene,
we have constructed maximally localized Wannier orbitals (WO's)~\cite{bib:MLWF} for the four conduction bands.
The DFT calculations were performed with 
the Tokyo Ab initio Program Package (TAPP)~\cite{bib:TAPP} 
and the Troullier-Martins norm-conserving pseudopotentials~\cite{bib:TM-NCPP}
within LDA.
The Bloch wave functions were expanded in terms of plane waves with an energy cutoff of 49 Ry.
The features of the calculated band structure of the conduction bands 
 (Fig.\ref{wannier} (b)), 
with the atomic positions optimized above, 
are identical to those obtained by QMAS within 
the numerical accuracy.  
Setting the energy window to cover the four conduction bands, 
we have obtained four maximally localized WO's, two of which 
($w_{\mathrm{A} l}, w_{\mathrm{A} h}$) are centered at molecule A 
while the other two
($w_{\mathrm{B} l}, w_{\mathrm{B} h}$) at molecule B.
The wave functions $w_{\mathrm{A} i}, w_{\mathrm{B} i}$ 
(Fig.\ref{wannier} (a)) have 
the same shape around each molecule for each of $i = l, h$ 
with $l(h)$ denoting  the lower (higher) energy states.
The band structure in Fig.\ref{wannier}(b) in the tight-binding 
model constructred from the WO's reproduces accurately the DFT-LDA band dispersion.  
The transfer integrals, 
$t_{ij}(\boldsymbol{d}) \equiv  \langle w_i(\boldsymbol{r}) | H_{\mathrm{KS}} | w_j(\boldsymbol{r} - \boldsymbol{d}) \rangle$, 
are defined as the matrix elements of the Kohn-Sham Hamiltonian $H_{\mathrm{KS}}$ 
between the WO's separated by $\boldsymbol{d} = d_1 \boldsymbol{a} + d_2 \boldsymbol{b} + d_3 \boldsymbol{c}$. 
The transfer integrals with significant magnitudes are depicted in Fig.\ref{wannier} (c).  
If we examine the shapes of 
$ w_l $ and $ w_h $, they can be represented approximately as 
$ \phi_{\mathrm{LUMO}}  \pm \phi_{\mathrm{LUMO} + 1} $, with 
$w_h$ having a slightly larger Wannier spread than that of $w_l$.
As seen in Fig.\ref{orbitals2} (a), the LUMO (LUMO$+1$) are 
antisymmetric (symmetric) with respect to the plane that 
bisects the long molecular axis, and 
$w_l$ and $w_h$ are nearly mirror images of each other, 
with a slight polarization from the center of the molecule.
The matrix element of $H_{\mathrm{KS}}$ between the LUMO and LUMO$+1$ in the crystal is approximately given by
$| \langle \phi_{\mathrm{LUMO}} | H_{\mathrm{KS}} | \phi_{\mathrm{LUMO}+1} \rangle | \simeq (\varepsilon_h - \varepsilon_l)/2 = 11.3$ meV,
where $\varepsilon_i \equiv t_{ii}(\boldsymbol{0})$.  
This is not negligible against the gap of 67.8 meV between the LUMO and LUMO$+1$ calculated for the molecule, which 
leads to the hybridization of the LUMO- and (LUMO$+1$)-derived bands.  
This contrasts with pentacene, for which the calculated gap in the molecule between the LUMO and LUMO$+1$ is as high as 1.28 eV, and the bands derived from them do not hybridize.

{\it Doped solid picene ---} 
Let us move on to the doped solid picene.  
In contrast to the case of BEDT-TTF or TMTSF,
carriers are electrons in K$_x$picene.  
If we assume rigid bands and shift the Fermi energy 
to a position corresponding to $x = 3$, 
CB3 and CB4 intersect the Fermi level, 
and the Fermi surface, displayed in Fig.\ref{band1} (b), 
shows a double surface structure~\cite{bib:1188}. 
However, the rigid band assumption has to be tested.  
The measured lattice parameters~\cite{kubozono} of the K$_{2.9}$picene are 
$a=8.707, b=5.912, c=12.97$ in \AA \ with $\beta=92.77^\circ$.  
One conspicuous feature is that 
the $c$ axis shrinks while the $a$ axis expands upon doping.  
This contrasts with the crystalline pentacene, where 
significant expansions of the $c$ axis are known to 
occur when doped with alkali elements~\cite{bib:alkali-doped_pentacene} 
as a result of the intercalants accommodated in the interlayer region.  
So this suggests that, in the case of picene, 
the dopants are not inserted into the interlayer region.

Another issure of interest is ``interlayer states" in GIC.  
There is a body of studies, both experimental~\cite{bib:koma} 
and theoretical~\cite{bib:1110}, that indicate 
the interlayer states of graphite that reside in the space between the 
carbon layers are relevant in the electronic structure in 
general and superconductivity in particular in GIC, 
where the interlayer states of the pristine graphite, located above the Fermi level, strongly hybridize with the valence states of the intercalant atoms
and the electronic bands responsible for the hybridized states descend to intersect the Fermi level.  
In solid picene, we have found that, while there exist interlayer states 
(where we define a layer as the array of molecules in the $ab$ plane) 
with amplitudes well away from the layers, they 
sit in energy 1.5 eV above the conduction band bottom (Fig.\ref{orbitals2} (b)).  
If the dopants are accommodated within the layer of the solid picene 
as suggested above, the dopant atoms are then expected to have 
little hybridization with the interlayer states of the solid picene.  

\begin{figure}[htbp]
\begin{center}
\includegraphics[keepaspectratio,width=6cm]{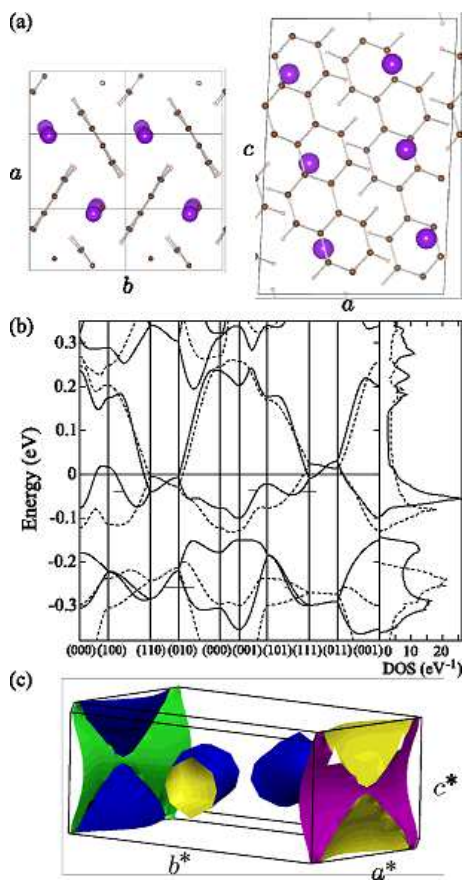}
\end{center}
\caption{
(a) Optimized K$_3$picene geometry with 
large balls representing K atoms.
(b) Band structure (solid curves with $E=0$ corresponding to $E_F$),
along with that for the K-removed geometry for comparison (dashed curves
with $E=0$ put equal to the position of $E_F$ when we shift it
to the situation for $x=3$).
(c) The Fermi surface for the K$_3$picene.
}
\label{band3}
\end{figure}

To determine whether intralayer insertion of K atoms is energetically more favorable than interlayer insertion,
we have prepared initial geometries by locating a K atom on each of $3 \times 3$ grid points on the $ab$ plane of the pristine picene,
from which we started optimization.  
The energies of the intralayer insertions are found to be 
about $0.2$ eV lower in energy per molecule than those of the interlayer insertions.  
We have therefore performed an optimization for K$_3$picene starting from an initial geometry where six K atoms are put in between the 
molecules (four above the end benzene rings 
and two above the center rings)  within the layer 
with the lattice parameters set to those observed for K$_{2.9}$picene.
The angle between the molecular planes obtained is now obtuse, $114^\circ$,
which is due to the expanded $a$ and shrunk $b$ axes.
Analysis of the resultant structure shows that the smaller volume of the unit cell and the presence of the dopants cause 
a larger distortion of the planar shape of the molecules than in the pristine case.
The electronic band structure is shown in Figure \ref{band3} (b) 
(solid lines).  
They significantly differ from those in the pristine picene, so that 
the rigid band assumption is invalid.  
In Fig.\ref{band3} (b) we have also displayed (dashed lines) 
the result for a geometry with the K atoms removed 
but the other atoms and the doping level fixed.  
While the two results differ to some extent, which implies that potassium-states 
are hybridized in the conduction bands,  
a stronger
origin of the non-rigidity against doping
comes from the difference in the molecular orientations, 
which dominates the overlaps of the $\pi$ MO's.
The Bloch states near the Fermi level, however, turn out 
to have still the character of the LUMO and LUMO$+1$, while 
interlayer states were not found near the Fermi level.

The Fermi surface for the doped system in Fig.\ref{band3} (c) 
consists of multiple surfaces, 
which contain warped, interwoven one-dimesional 
(planar) surfaces and 
a three-dimensional (pocket) surface, some of which involve 
transfers between MO's via K atoms.

To summarize we have obtained the band structure of 
solid picene with curious features.  The geometry optimized here is not a decisive one, since there can be many metastable geometries, which was also the case with solid pentacene~\cite{bib:1095}, 
and further investigations should be needed.  
Future problems include how  the features of the Fermi surface 
will have implications on  mechanisms of 
superconductivity.

We are indebted to Yoshihiro 
Kubozono for letting us know of the experimental 
results prior to publication.  
The present work is partially supported by the Next Generation Supercomputer Project,
Nanoscience Program from MEXT, Japan, and 
by Scientific Research on Priority Areas  "New Materials Science Using Regulated Nano Spaces"
under Grant No. 19051016 from MEXT, Japan.
The calculations were performed on the AIST Super Cluster at the Tsukuba Advanced Computing Center (TACC), AIST.


\begin{thebibliography}{99}

\bibitem{bib:1114}
N. B. Hannay, T. H. Geballe, B. T. Matthias, K. Andres, P. Schmidt and D. MacNair,
Phys. Rev. Lett. {\bf 14} (1965) 225.

\bibitem{bib:1108_10-11}
T. E. Weller, M. Ellerby, S. S. Saxena, R. P. Smith and N. T. Skipper, 
Nature Phys. {\bf 1} (2005) 39; 
N. Emery, C. H\'erold, M. d'Astuto, V. Garcia, Ch. Bellin, J. F. Mar\^ech\'e, P. Lagrange and G. Loupias, 
Phys. Rev. Lett. {\bf 95} (2005) 087003.

\bibitem{bib:1117_9}
D. Jerome, A. Mazaud, M. Ribault and K. Bechgaard, 
J. Phys. Lett. (Paris) {\bf 41} (1980) L95.

\bibitem{bib:1108_13}
H. Taniguchi, M. Miyashita, K. Uchiyama, K. Satoh, N. Mori, H. Okamoto, K. Miyagawa, K. Kanoda, M. Hedo and Y. Uwatoko, 
J. Phys. Soc. Jpn. {\bf 72} (2003), 468.

\bibitem{bib:1104_1}
A. F. Hebard, M. J. Rosseinsky, R. C. Haddon, D. W. Murphy, S. H. Glarum, T. T. M. Palstra, A. P. Ramirez and A. R. Kortan, 
Nature {\bf 350} (1991) 600.

\bibitem{bib:1104_4}
K. Tanigaki, T. W. Ebbesen, S. Saito, J. Mizuki, J. S. Tsai, Y. Kubo and S. Kuroshima,
Nature {\bf 352} (1991) 222.

\bibitem{bib:1104_8}
T. T. M. Palstra, O. Zhou, Y. Iwasa, P. E. Sulewski, R. M. Fleming and B. R. Zegarski, 
Solid State Commun. {\bf 93} (1995) 327.

\bibitem{kubozono}
R. Mitsuhashi, Y. Suzuki, Y. Yamanari, T. Kambe, N. Ikeda, H. Okamoto, 
A. Fujiwara, M. Yamaji, N. Kawasaki, Y. Maniwa, and 
Y. Kubozono, private communications.

\bibitem{bib:1106}
H. Okamoto, N. Kawasaki, Y. Kaji, Y. Kubozono, A. Fujiwara and M. Yamaji, 
J. Am. Chem. Soc. {\bf 130} (2008) 10470.

\bibitem{bib:1107}
N. Kawasaki, Y. Kubozono, H. Okamoto, A. Fujiwara and M. Yamaji, 
Appl. Phys. Lett. {\bf 94} (2009) 043310.

\bibitem{bib:1079}
A. De, R. Ghosh, S. Roychowdhury and P. Roychowdhury, 
Acta Cryst. C {\bf 41} (1985) 907.

\bibitem{bib:PAW}
P. E. Bl\"ochl, Phys. Rev. B {\bf 50} (1994) 17953; 
G. Kresse and D. Joubert, {\it ibid.} {\bf 59} (1999) 1758.

\bibitem{bib:QMAS}
http://www.qmas.jp/


\bibitem{bib:1087}
K. Hummer and C. Ambrosch-Draxl, Phys. Rev. B {\bf 72} (2005), 205205.

\bibitem{bib:MLWF}
N. Marzari and D. Vanderbilt, Phys. Rev. B {\bf 56} (1997) 12847;
I. Souza, N. Marzari and D. Vanderbilt, Phys. Rev. B {\bf 65} (2001) 035109.

\bibitem{bib:TAPP}
J. Yamauchi, M. Tsukada, S. Watanabe and O. Sugino, 
Phys. Rev. B {\bf 54} (1996) 5586.

\bibitem{bib:TM-NCPP}
N. Troullier and J. L. Martins, Phys. Rev. B {\bf 43} (1991) 1993.

\bibitem{bib:1188} 
Relevance of multiple LUMO's is reminiscent of Ni(tmdt)$_2$ 
[H. Seo, S. Ishibashi, Y. Okano, H. Kobayashi, A. Kobayashi, H. Fukuyama and K. Terakura, J. Phys. Soc. Jpn. {\bf 77} (2008) 023714], 
but this system is a single-component $\pi$-$d$ system.

\bibitem{bib:alkali-doped_pentacene}
T. Minakata, I. Nagoya and M. Ozaki, 
J. Appl. Phys. {\bf 69} (1991) 7354;
T. Minakata, H. Imai and M. Ozaki, J. Appl. Phys. {\bf 72} (1992) 4178; 
T. Ito, T. Mitani, T. Takenobu and Y. Iwasa, 
J. Phys. Chem. Solids {\bf 65} (2004) 609;
Y. Matsuo, T. Suzuki, Y. Yokoi and S. Ikehata, 
J. Phys. Chem. Solids {\bf 65} (2004) 619;
Y. Matsuo, S. Sakai and S. Ikehata, 
Phys. Lett. A {\bf 321} (2004) 62;
B. Fang, H. Zhou and I. Homma, 
Appl. Phys. Lett. {\bf 86} (2005) 261909.

\bibitem{bib:koma}
A. Koma, K. Miki, H. Suematsu, T. Ohno and H. Kamimura, 
Phys. Rev. B {\bf 34} (1986) 2434.

\bibitem{bib:1110}
G. Cs\'anyi, P. B. Littlewood, A. H. Nevidomskyy, C. J. Pickard and B. D. Simons, 
Nature Phys. {\bf 1} (2005) 42.

\bibitem{bib:1095}
R. G. D. Valle, E. Venuti and A. Brillante, 
J. Chem. Phys. {\bf 118} (2003) 807.


\end{thebibliography}
\end{document}